\documentclass{IEEEjmw}

\usepackage[colorlinks,urlcolor=blue,linkcolor=blue,citecolor=blue]{hyperref}

\usepackage{color,array}

\usepackage{graphicx}

\usepackage{textcomp}
\usepackage{stfloats}
\usepackage{url}
\usepackage{verbatim}
\usepackage{amsmath,amsfonts}
\usepackage{algorithmic}
\usepackage{algorithm}
\usepackage{array}
\usepackage{siunitx}
\usepackage{cite}

\jvol{00}
\jnum{XX}
\paper{1234567}
\pubyear{20XX}
\receiveddate{Month~XX, 20XX}
\accepteddate{Month~XX, 20XX}
\publisheddate{Month~XX, 20XX}
\currentdate{Month~XX, 20XX}
\doiinfo{JMW.20XX.Doi Number}

\begin{document}

\sptitle{Article Category}

\title{Super-Resolution Radar Imaging with Sparse Arrays Using a Deep Neural Network Trained with Enhanced Virtual Data} 

\editor{The associate editor coordinating the review of this manuscript and approving it for publication\break was F. A. Author.}

\author{CHRISTIAN SCHUESSLER\affilmark{1} (Graduate Student Member, IEEE)}

\author{MARCEL HOFFMANN\affilmark{1}  (Graduate Student Member, IEEE)}

\author{AND MARTIN VOSSIEK\affilmark{1} (Fellow, IEEE)}

\affil{Institute of Microwaves and Photonics, Friedrich-Alexander-Universität Erlangen-Nürnberg, Erlangen, Germany}

\corresp{CORRESPONDING AUTHOR: Christian Schuessler (e-mail: \href{mailto:christian.schueslser@fau.de}{christian.schuessler@fau.de}).}

\markboth{PREPARATION OF PAPERS FOR IEEE JOURNAL OF MICROWAVES}{F. A. AUTHOR {\itshape ET AL}}

\begin{abstract}
    This paper introduces a method based on a deep neural network (DNN) that is perfectly capable of processing radar data from extremely thinned radar apertures.
    The proposed DNN processing can provide both aliasing-free radar imaging and super-resolution. The results are validated by measuring the detection performance on realistic 
    simulation data and by evaluating the Point-Spread-function (PSF) and the target-separation performance on measured point-like targets. Also, a 
    qualitative evaluation of a typical automotive scene is conducted.     
    It is shown that this approach can outperform state-of-the-art subspace algorithms and also other existing machine learning solutions. 
    The presented results suggest that machine learning approaches trained with sufficiently sophisticated virtual input data are a
    very promising alternative to compressed sensing and subspace approaches in radar signal processing.
    The key to this performance is that the DNN is trained using realistic 
    simulation data that perfectly mimic a given sparse antenna radar array hardware as the input. 
    As ground truth, ultra-high resolution data from an enhanced virtual radar are simulated. 
    Contrary to other work, the DNN utilizes the complete radar cube and not only the antenna channel information at certain range-Doppler detections.
    After training, the proposed DNN is capable of sidelobe- and ambiguity-free imaging. It simultaneously delivers 
    nearly the same resolution and image quality as would be achieved with a fully occupied array. 
\end{abstract}


\maketitle

\section{INTRODUCTION}
Radar sensors are among the most popular sensors for automotive driving applications. Compared to lidar and camera sensors, 
they are robust, cheap, and do not suffer due to 
harsh weather conditions, such as snow, rain, and fog~\cite{lit:rosique2019systematic}.
Also, radar sensors can directly measure the radial speed of objects by exploiting the Doppler effect, 
which drastically improves the classification of moving entities such as bicycles and pedestrians~\cite{lit:prophet2018pedestrian,lit:heuel2012pedestrian,lit:palffy2020cnn}.

However, one of the main disadvantages of radar sensors is the low angular resolution, which is typically below that of lidar sensors. 

\begin{figure}
	\centerline{\includegraphics[width=0.48\textwidth]{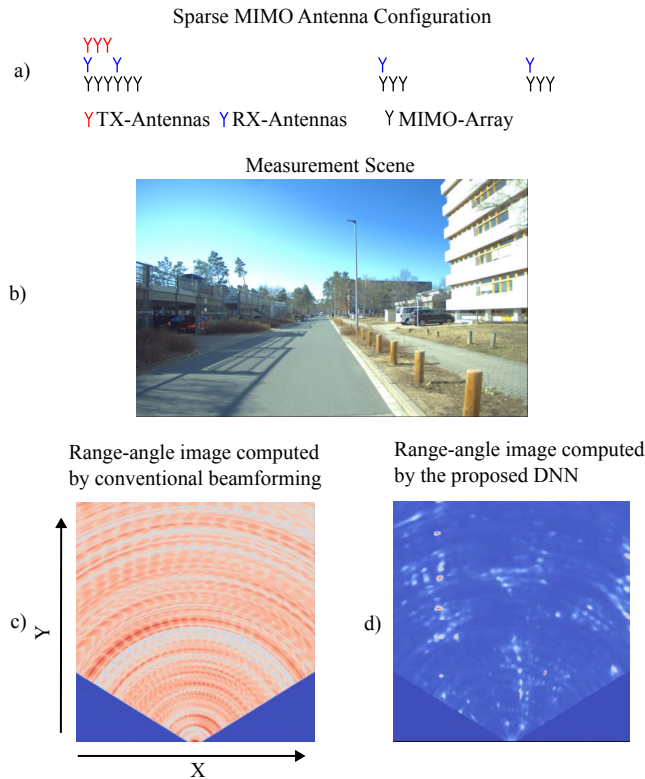}}
	\caption{Image a{)} shows the sparse MIMO antenna arrray configuration 
        that is derived from a small subset of the antennas of the AVR-QDM-110 radar system from Symeo GmbH, an indie Semiconductor company.
        The measurement scene used for evaluation is depicted in b{)}.
        The results for the conventional delay-and-sum (DaS) beamforming and the proposed approach 
        using a deep neural network (DNN) applied to the sparse 
        array configuration are presented in c{)} and d{)}, respectively.
        As visible the DaS beamformer is not able to process the sparse data at all, 
        wheras the DDN trained with our novel approach yield a nearly perfect result.}
	\label{fig:FirstFigure}
\end{figure}

Therefore, the aim of many hardware and signal processing approaches is to obtain high-resolution angular information.
To limit both the hardware complexity and the amount of received data, one objective is to restrict the number of physical antennas.
Among the most common ways to achieve this is by using multiple-input multiple-output (MIMO) antenna arrays,
followed by subspace methods for direction of arrival (DoA) estimation, such as 
MUSIC~\cite{lit:schmidt1986multiple}, ESPRIT~\cite{lit:roy1989esprit}, or, more recently,
compressed sensing (CS) methods~\cite{lit:strohmer2009compressed, lit:roos2019compressed, lit:correas2018experimental, lit:alistarh2022compressed, lit:hu2023off, lit:de2019compressed}.

Another approach to reduce the number of antennas required is to sparsely sample the entire array aperture 
by not spacing the antennas equidistantly. This is done, for example, by deploying minimum redundancy arrays~\cite{lit:moffet1968minimum}, 
co-prime arrays~\cite{lit:vaidyanathan2010sparse}, or nested arrays~\cite{lit:pal2010nested}.
A further approach is to design the antenna placement by optimizing certain properties of the expected signal, 
for example by minimizing the mainlobe width and at the same time constraining or minimizing the sidelobe level as demonstrated
in~\cite{lit:pavlenko2017design} and~\cite{lit:mateos2019sparse}, or more suitable for CS applications, 
the coherence of the signal~\cite{lit:eisele2022novel}.

In this work, high-resolution and aliasing-free radar images are generated from sparse antenna configurations that would 
lead to severe ambiguities and aliasing with conventional signal processing methods.
The proposed deep learning–based signal processing scheme is not only applied to simulated data 
but also tested on real measurements, as shown in Fig.~\ref{fig:FirstFigure}.

In our previous work~\cite{lit:schussler2022deep}, we applied a deep neural network (DNN) to radar range-angle images 
to drastically alleviate clutter and noise in real-world measurement scenarios. 
This was achieved by simulating both enhanced high-resolution noise and clutter-free data as the ground truth 
and radar data of an accurate digital twin of the radar sensor under test as input. 
As a consequence, the DNN, which was trained exclusively in a simulation environment, 
successfully learned to remove clutter and sharpen the image.
The simulations were generated using the radar ray tracing implementation we proposed in~\cite{lit:schussler2021realistic}.

Compared to other existing work, our previous work in~\cite{lit:schussler2022deep} firstly utilized realistic radar simulations for radar image enhancement 
with respect of noise and clutter reduction. 
There and in this work, the simulations were conducted in a realistic environment, resembling a city landscape, see Fig.~\ref{fig:ExampleImages} (a).
This way, real-world structures, such as building, trees, and other objects are accurately mapped to the generated radar images.
However, this was implemented for a fully occupied array and only the magnitude of the simulated 
range-angle image was used as input.

Here, our previous work is extended in two fundamental ways.
Firstly, we consider severely thinned antenna arrays and show that 
even for extremely sparse array configurations, an image reconstruction is still possible. 
Secondly, a novel feature vector is proposed, 
which combines information by applying beamsteering on the sparse antenna array and utilizing the 
covariance matrix from raw channel data at the same time. 
Further, the training data is extended by synthetic data with closely spaced targets to improve the target separation performance 
even further. 
Compared to our previous work, an algorithm to extend the approach for multiple Doppler detections 
in the same range bin is proposed and evaluated.
Also, the antenna radiation pattern is considered in the training data simulation, which was omitted in our previous work.

Compared to other existing work for DoA estimation, this work firstly utilizes the complete range-angle-image selected at specific Doppler detections instead of 
using only single range-Doppler detections. 
Also, the usage of high realistic simulation data for DoA estimation, especially for sparse arrays, is completely novel. 

\begin{figure}
	\centerline{\includegraphics[width=0.49\textwidth]{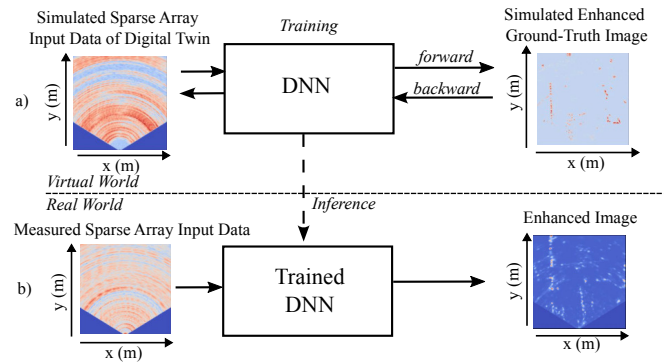}}
	\caption{The image shows the overall concept of this approach. In a{)},
    the complete DNN is trained using only simulated data and attempts to reproduce the high-resolution simulated 
    and enhanced ground truth data as well as possible.
    After training, the DNN can be applied in 
    real measurement scenarios (inference), as shown in b{)}.}
	\label{fig:Concept}
\end{figure}

In detail, for the training of the DNN, two special datasets are generated with our realistic ray tracing simulator presented in~\cite{lit:schussler2021realistic}.
The first simulated dataset is designed to resemble realistic measurement data as close as possible, serving as a digital twin of the real-world sensor.
The second dataset is simulated using the corresponding enhanced virtual sensor with the same parameters but with a significant higher number of antennas, leading to a higher angular resolution.
The proposed training concept is briefly summarized in Fig.~\ref{fig:Concept}.

Consequently, the first dataset can be used as the input for the DNN, whereas the second dataset can be used as ground truth. This will be further described in Section~\ref{sec:SigProc}.
Because of the easy scalability of simulation data, the DNN can be trained very thoroughly with large input and ground truth datasets.
Since the simulated data is so close to reality, this DNN can then be directly applied to real measurement data.

The training and preprocessing code of the proposed approach is also published here: 
\href{https://github.com/ChristianSchuessler/Sparse-Array-Radar-Imaging}{https://github.com/ChristianSchuessler/Sparse-Array-Radar-Imaging}.

\section{RELATED WORK}\label{sec:RelatedWork}
This section provides an overview of the existing work on sparse array processing and deep learning approaches for radar imaging.

In~\cite{lit:friedlander1990direction,lit:hott2018joint,lit:sim2019enhanced},
signal processing techniques were presented to interpolate the signal between antenna channels of sparse array configurations 
or to extrapolate small antenna arrays to larger ones.
Recently, machine learning algorithms were applied to the array interpolation or extrapolation problem in~\cite{lit:orr2021coherent},~\cite{lit:eschbaumer2022analysis}, and~\cite{lit:roldan2023self}.
In~\cite{lit:orr2021coherent}, this was done by using real measurement data in a self-supervised way. 
For the input data, certain channels of the fully occupied array were ignored and the remaining
channels have to be predicted by the neural network. 
This work uses a similar network architecture (U-Net) as our approach, 
but our work does not require full range-Doppler-channel tensors, instead only 2D images as input turned out be sufficient. Further, it directly predicts 
the reconstructed image by realistic, but high resolution simulated ground-truth data. This ground-truth data is completely noise free 
and has a much higher resolution compared to the deployed fully occupied array, which cannot be achieved by common self-supervised approaches based on real measurement data.

In comparison,~\cite{lit:eschbaumer2022analysis} was purely based on synthetic data in combination 
with a smaller neural network for training.
The authors suggest that interpolating the signal requires neural networks with fewer parameters compared to estimating the DoA directly.
Interestingly, even the target separation performance could be improved, the results seem to be to some extend unstable 
for single target scenarios at different bearing angles. 

The DoA can also be estimated directly by neural networks, as shown in various works, 
such as~\cite{lit:fuchs2022machine, lit:ma2022deep, lit:huang2018deep, lit:papageorgiou2021deep, 
lit:gall2020learning, lit:sang2021doa, lit:liu2018direction}.
These methods typically outperform traditional methods, such as MUSIC or ESPRIT, 
especially in environments with a low signal-to-noise ratio (SNR)~\cite{lit:papageorgiou2021deep}.
Most works in this context have relied on training data from synthetic point-like targets, 
requiring only a single snapshot of the antenna channels. However, in~\cite{lit:sang2021doa},
real measurement data and a region of interest in the fast-time were additionally 
used as input for the neural network.
In~\cite{lit:pavel2021machine} and~\cite{lit:pavel2022neural}, DoA approaches were also applied to 
distributed sparse arrays. 
The performance of our approach is compared to~\cite{lit:papageorgiou2021deep} in the results and shows a 
significant better performance, due to more realistic training data and by utilizing complete range-angle images.

Similar approaches can be found in the radar imaging domain, where the DoA of a signal is only considered implicitly, 
since a complete image has to be reconstructed at once. 
The authors in~\cite{lit:cheng2020compressive} used CS reconstructions of fully sampled simulated and 
measured data to teach the network to reproduce these results from low-resolution data. 
They consequently called their approach \emph{CS-DNN}, since it can produce similar or even better 
results compared to traditional CS methods.
Other similar approaches using DNNs in the radar imaging domain 
can be found in~\cite{lit:gao2018enhanced, lit:dai2020imaging,lit:salucci2022artificial,lit:rostami2022deep, lit:wei2018deep, lit:xu2020deep}.
Our approach adopts the concept of using 2D images as input by utilizing range-angle images, which are suited for automotive radar applications.

The reason for the popularity of NN architectures is their often better performance compared to traditional subspace or compressed sensing methods.
Also, typically most DoA NN architectures do not need an estimation of the number of targets, e.g.~model order as compared to subspace methods such as MUSIC or ESPRIT.
For CS methods, the number of targets does not have to be estimated directly, but also hyperparameters for different loss terms and the number of iterations have to be tuned.
However, tasks like proper feature engineering, collecting and generating enough training data can make NN architectures hard to build. 

\section{RADAR SYSTEM SETUP AND SIGNAL PROCESSING}\label{sec:SigProc}

In this section, the radar system, the frequency modulated continuous wave 
(FMCW) modulation scheme, and the corresponding signal processing workflow are presented. 
Following that, the input and ground truth data generation for the DNN is described in more detail.

Here, matrices are denoted in upper-case bold letters ($\mathbf{M}$), 
vectors in lower-case bold letters ($\mathbf{v}$), and scalars in lower-case letters ($a$).

\subsection{RADAR SYSTEM SETUP}

\begin{figure}
	\centerline{\includegraphics[width=0.3\textwidth]{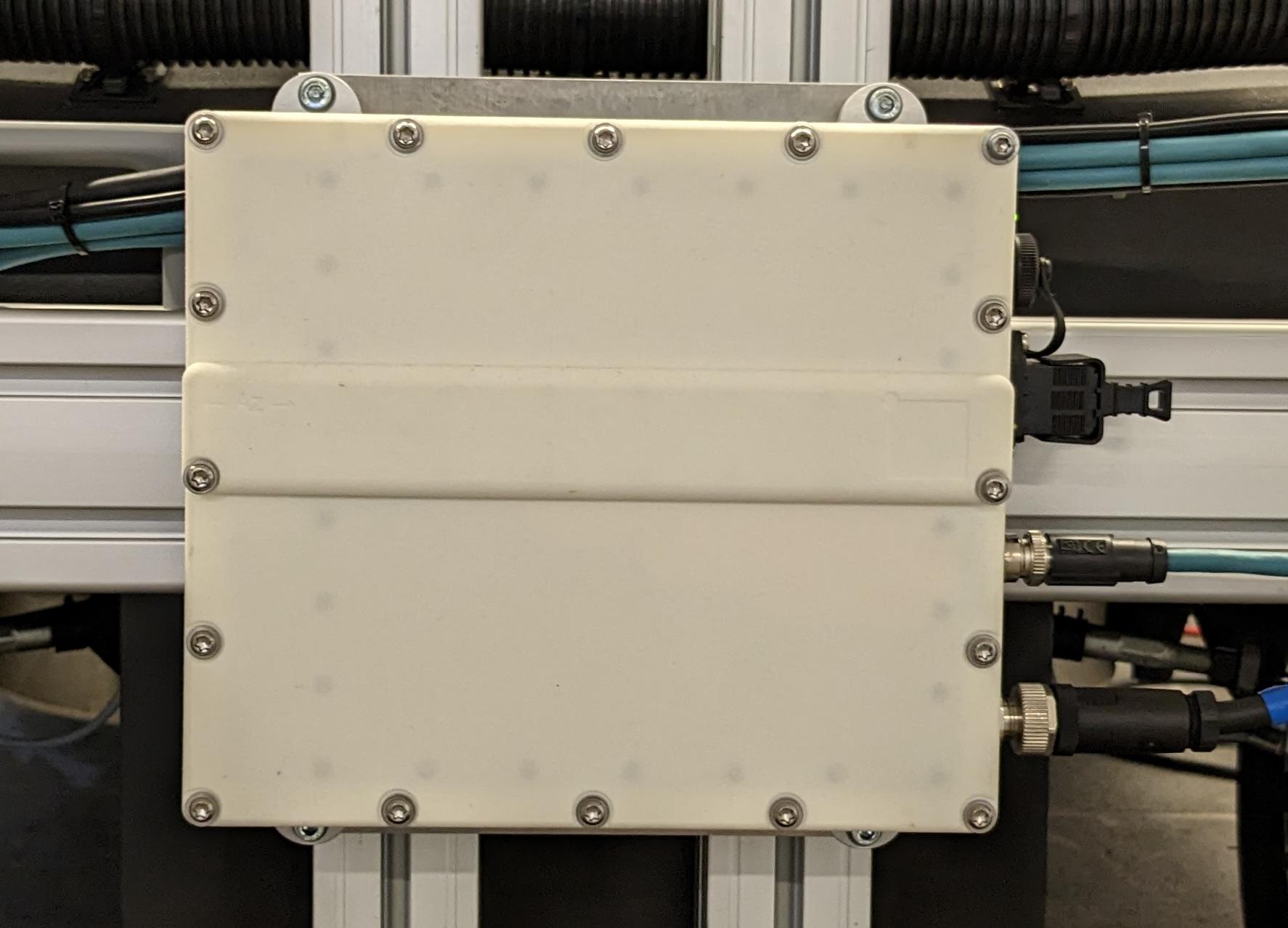}}
	\caption{Image of the deployed automotive FMCW MIMO radar 
    device AVR-QDM-110 from Symeo GmbH, an indie Semiconductor company.}
	\label{fig:RadarPhoto}
\end{figure}

\begin{figure}
	\centerline{\includegraphics[width=0.5\textwidth]{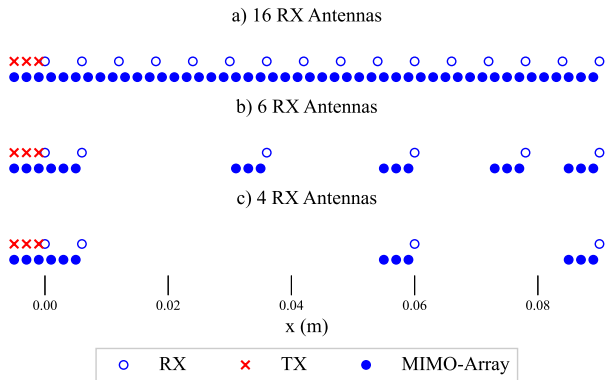}}
	\caption{The full antenna array of the chosen array configuration of the AVR-QDM-110 
    with 3 TX and 16 RX antennas, and consequently 48 virtual elements in the MIMO array, is shown in a{)} with 
    an virtual element spacing of 0.58$\lambda$. Two sparse variants are formed by omitting several RX antennas, in b{)} and c{)}.}
	\label{fig:AntennaConfig}
\end{figure}

An image of the radar unit attached to a test vehicle is depicted in Fig.~\ref{fig:RadarPhoto}.
The radar system consists of a MIMO FMCW radar device with the parameters stated in Table~\ref{tab:RadarParameters}.
Since the radar operates in MIMO mode, a virtual array can be constructed 
by combining the TX and RX antenna channels~\cite{lit:sun2020mimo}.
Sparse virtual arrays can be constructed by omitting certain TX and RX channels in the subsequent processing
chain. Certain antenna configurations with their corresponding virtual arrays, which are evaluated 
in this work, are depicted in Fig.~\ref{fig:AntennaConfig}. 
As can be seen, while the fully occupied array can be constructed well with the MIMO configuration of the 
investigated radar sensor, it is difficult to construct meaningful sparse array configurations.
Therefore, we tried to keep the overall aperture size large, while 
restricting the number of RX antennas drastically. 
Fig.~\ref{fig:AntennaConfig} shows that there are still pairs of three virtual antennas, which is due to the 
original MIMO configuration of the radar sensor. However, this article shows that even under this
suboptimal configuration high resolution ambiguity-free radar images can be obtained.

{
\setlength{\tabcolsep}{1mm}%
\newcommand{\CPcolumnonewidth}{12mm}%
\newcommand{\CPcolumntwowidth}{31mm}%
\newcommand{\CPcolumnthirdwidth}{18mm}
\newcommand{\CPcell}[1]{\hspace{0mm}\rule[-0.3em]{0mm}{1.3em}#1}%
\newcommand{\CPcellbox}[1]{\parbox{90mm}{\hspace{0mm}\rule[-0.3em]{0mm}{1.3em}#1\strut}}%
\begin{table}
\caption{Radar Parameters}
\small
\centering
\begin{tabular}{|l|l|l|}\hline
\parbox{\CPcolumnonewidth}{\CPcell{\bfseries Symbol}} & \parbox{\CPcolumntwowidth}{\CPcell{\bfseries Parameter}} & \parbox{\CPcolumnthirdwidth}{{\CPcell{\bfseries Value}}} \\ \hline
\CPcell{$f_\mathrm{c}$ } & \CPcell{RF carrier frequency} & \CPcell{\SI{77}{GHz}}  \\ \hline
\CPcell{$B$} & \CPcell{RF bandwidth} & \CPcell{\SI{1}{GHz}} \\ \hline
\CPcell{$N_{\mathrm{chirp}}$} & \CPcell{number of chirps} & \CPcell{128} \\ \hline
\CPcell{$T_\mathrm{c}$} & \CPcell{Chirp duration} & \CPcell{\SI{80.6}{\mathrm{\mathrm{\mu} s}}}\\ \hline
\CPcell{$N_{\mathrm{TX}}$} & \CPcell{Number of TX antennas} & \CPcell{3}\\ \hline 
\CPcell{$N_{\mathrm{RX}}$} & \CPcell{Number of RX antennas} & \CPcell{16} \\ \hline
\CPcell{$d_{\mathrm{TX}}$} & \CPcell{TX antenna element distance} & \CPcell{\SI{2}{mm}} \\ \hline
\CPcell{$d_{\mathrm{RX}}$} & \CPcell{RX antenna element distance} & \CPcell{\SI{6}{mm}} \\ \hline
\CPcell{$\delta_\mathrm{a}$} & \CPcell{Angular resolution} & \CPcell{\SI{1.83} deg.} \\ \hline
\end{tabular}
\label{tab:RadarParameters}
\end{table}
}

\subsection{FMCW SIGNAL MODEL}

The intermediate frequency (IF) signal, also called beat signal, of an FMCW radar can be given as

\begin{equation}\label{equ:ifsignal}
    \begin{split}
& s_{\mathrm{IF}}(i_{\textrm{TX}}, i_{\textrm{RX}}, t) = \\
& \sum_{k=0}^{N-1} A_k \mathrm{\exp}(2 \pi \mathrm{j} (\mu t \tau_k(i_{\textrm{TX}}, i_{\textrm{RX}}) + f_\textrm{c} \tau_k( i_{\textrm{TX}}, i_{\textrm{RX}})))\textrm{,}\\
\end{split}
\end{equation}
with each transmit (TX) and receive (RX) antenna channel index denoted 
as $i_{\mathrm{TX}}$ and $i_{\mathrm{RX}}$, respectively. The complete signal is composed as a coherent summation
of the modulation term including the round trip delay $\tau_k$ 
of all targets $N$ in the scene with amplitude $A_k$, as defined in the equation below:
\begin{equation}\label{equ:roundtrip}
    \tau_k(i_{\mathrm{TX}},i_{\mathrm{RX}}) = \frac{||\mathbf{x}_{i_{\textrm{TX}}} - \mathbf{x}_k||_2 + ||\mathbf{x}_{i_{\textrm{RX}}} - \mathbf{x}_k||_2}{c} \textrm{,}
\end{equation}
with $c$ being the speed of light and $\mathbf{x}_{\textrm{TX}}$, $\mathbf{x}_{\textrm{RX}}$, and $\mathbf{x}_k$ 
representing the position of the antennas and targets, respectively.
The frequency slope $\mu$, also called the chirp-rate of the FMCW radar, is 
given by the bandwidth $B$ divided by the chirp duration $T_c$

\begin{equation}\label{equ:frequencyslop}
    \mu = \frac{B}{T_\textrm{c}} \textrm{.}
\end{equation}
A more detailed description of FMCW radar signal processing can be found in 
various review and tutorial papers, such as~\cite{lit:li2021signal}.

For the ground truth data simulation, the identical radar parameters to those described in Table~\ref{tab:RadarParameters} are used. 
However, in contrast to 
the sparse input data generation, a single TX is combined with 256 RX antennas that are placed as a uniform linear array (ULA).
The 3-dB beamwidth (i.e.~resolution) for linear arrays can be estimated as~\cite[p.~11]{lit:richards2014fundamentals} 
\begin{equation}
\theta_3 = \frac{51.05 \lambda}{D},
\end{equation}
for wavelength $\lambda$ and aperture length $D$.
The resolution of the full not sparse input antenna array is then:
\begin{equation}
\theta_3^i \approx \frac{51.05 \lambda}{48 \cdot 0.58 \lambda} = 1.83 \textrm{ deg.},
\end{equation}   
and for the enhanced ground truth radar sensor
\begin{equation}
\theta_3^g \approx \frac{51.05 \lambda}{256 \cdot 0.5 \lambda} = 0.4 \textrm{ deg.}.
\end{equation}

\subsection{INPUT DATA PROCESSING}\label{subsec:InputData}
For all the following computations, we assume far-field conditions, meaning that spherical waves originating from point-scatterers can be approximated by plane waves, 
given that the scatterer is sufficiently far away. 
This assumption implies that each target lies in the same range bin, 
which is typical in automotive radar applications~\cite{lit:sun2020mimo}.

First, an FFT is applied along the fast- and slow-time dimension of $s_{\textrm{IF}}$ to 
obtain range and Doppler information for each TX–RX channel combination. In the next step only the magnitude of all 
range-Doppler images is taken to be processed into a single \emph{mean} image using all range-Doppler images. 
This does improve the SNR for the following Doppler bin selection, and also ensures that 
for each antenna channel the same range-Doppler pixel is taken in all further processing steps.

Afterwards, one single Doppler bin, typically the bin with maximum amplitude, is selected for each range bin.
This leads to the 2D signal $S_{\textrm{IF}}$ in the frequency domain, that has dimensions in channel and range. 
In the second stage, the TX and RX antenna channels are sorted to form a virtual antenna array, 
which corresponds to a non-uniform linear array, as in Fig.~\ref{fig:AntennaConfig} b{)} and c{)}.

Essentially, two features are generated out of the signal $S_{\textrm{IF}}$ as described below:
\begin{enumerate}
    \item Typically, for ULA configurations, a fast fourier transform (FFT) is also applied in the channel direction to obtain angular information.
    However, this is not immediately possible for sparse arrays, since 
    there are gaps in the signal. Instead, a common delay-and-sum (DaS) 
    beamforming method that applies steering weights for an angle hypothesis is chosen~\cite{lit:richards2014fundamentals}.
    The steering weights for an arbitrary array configuration are shown below:
    
    \begin{equation}\label{equ:steeringvec}
        \mathbf{v}(\theta) = 
        \begin{pmatrix}
        w_0 \cdot e^{- \textrm{j} \frac{2\pi}{\lambda} \sin{(\theta)} \mathbf{x}_0} \\
        \vdots\\
        w_{N_\textrm{v}-1} \cdot e^{-\textrm{j} \frac{2\pi}{\lambda} \sin{(\theta)} \mathbf{x}_{N_\textrm{v}-1}} \\
        \end{pmatrix} \textrm{,}
    \end{equation}
    where $\mathbf{x}_i$ denotes the one dimensional virtual array position, 
    and $\theta$ is the current steering angle. 
    The number of virtual antennas is denoted by $N_\mathrm{v}$, which is the product of $N_\mathrm{TX}$ and $N_\mathrm{RX}$.
    To reduce the sidelobe level, a window function implemented via the weights $w$ can be applied. 
    However, for simplicity all weights $w$ are set to 1, since this approach already lead to good results.
    Moreover, we apply the beamsteering in the sine domain, such as~for $u=\sin(\theta)$,
    so that this approach is also suited to extrapolate linear arrays 
    where the beamsteering processing can be replaced by an often faster FFT.
    
    The complete beamforming process for all angles $\theta$, resulting in the vector $\mathbf{i}_s$,
    can then be stated as a matrix multiplication for each range index $r_i$ 
    with the processed signal $\mathbf{S}_{\textrm{IF}}$ with steering matrix 
    $\mathbf{V}$, as shown below:
    
    \begin{equation}\label{equ:beamsteering}
        \mathbf{i}_s(r_i) = \mathbf{S}_{\textrm{IF}}(r_i) \mathbf{V}\textrm{.}
    \end{equation}
    
    The shape of $\mathbf{V}$ is $N_\mathrm{v} \times N_{\theta}$, with $N_{\theta}$ being the number of steering angles.
    The entries of $\mathbf{V}$ are
    \begin{equation}\label{equ:beamsteeringMatrix}
        \mathbf{V} = 
        \begin{pmatrix}
        e^{- \textrm{j} \frac{2\pi}{\lambda} u_0 x_0}  & \hdots &  e^{- \textrm{j} \frac{2\pi}{\lambda} u_{N_\theta} x_0} \\
        \vdots & \ddots & \vdots\\
        e^{- \textrm{j} \frac{2\pi}{\lambda} u_0 x_{N_\textrm{v}-1}} & \hdots & e^{- \textrm{j} \frac{2\pi}{\lambda} u_{N_\theta} x_{N_\textrm{v}-1}}\\
        \end{pmatrix}.
    \end{equation}
    The computed vector $\mathbf{i}_s(r_i)$ has the length $N_\theta$, which is equal to the required angular bins for the DNN. 
    For ULA configurations, this matrix multiplication is equal to a discrete 
    Fourier transform~\cite{lit:richards2014fundamentals} and can therefore be replaced by an FFT.
    This feature was selected, because it already worked very well in our previous work for radar image enhancement~\cite{lit:schussler2022deep}.
    Instead of only using the magnitude, additional phase information of the signal was selected as feature in this work, as it showed good results 
    in~\cite{lit:cheng2020compressive}.

    \item Secondly, the covariance matrix is used as an input for the DNN.
    It has a shape of $N_\mathrm{v} \times N_\mathrm{v}$ and is computed for each range bin as described in the equation below:
    \begin{equation}\label{equ:DNNCovMat}
        \mathbf{\Sigma}(r_i) = \frac{\mathbf{S}_{\mathrm{IF}}(r_i) \mathbf{S}_{\mathrm{IF}}^{\mathrm{H}}(r_i)}{||\mathbf{S}_{\mathrm{IF}}(r_i) \mathbf{S}_{\mathrm{IF}}^\mathrm{H}(r_i)||_2}.
    \end{equation}
    The symbol $|| \cdot ||_2$ denotes the Frobenius norm of the matrix.
    Since the resulting matrix is Hermitian, only the upper or lower triangular is utilized for further processing, 
    as already shown by~\cite{lit:liu2018direction}. This approach saves memory
    for larger antenna arrays, since our network is only designed for 450 angular bins. 
    Furthermore, in order to fit a single angular line for each range index,
    the covariance matrix has to be unrolled to a vector and padded with zeros in case 
    the resulting vector is smaller than the number of angular bins.
    The complete operation is summarized in the equation below:
    \begin{equation}\label{equ:CovMatunroll}
        \mathbf{i}_{\mathrm{cov}}(r_i) = \mathrm{pad}(\mathrm{unroll}(\mathrm{triu}({\mathbf{\Sigma}(r_i)}))).\\
    \end{equation}
    For large antenna arrays (e.g.,~full array), the resulting vector would be too large for the designed angular bins.
    In this case, the matrix is cropped symmetrically before unrolling. This discards some information 
    but still improves the results considerably.

    The real and imaginary part and as well as the phase information of the covariance was chosen 
    since it lead to good results in the work~\cite{lit:papageorgiou2021deep} for DoA estimation.
\end{enumerate}

The final input matrix $\mathbf{I}_{\mathrm{in}}(r_i)$ is real valued and composed of the following $N_{\mathrm{feat}}$ features:
\begin{equation}\label{equ:InputDNN}
    \mathbf{I}_{\mathrm{in}}(r_i) =  
        \begin{pmatrix}
            \mathrm{Abs}({\mathbf{i}_\mathrm{s}(r_i)})\\
            \mathrm{Phase}(\mathbf{i}_\mathrm{s}(r_i))\\
            \mathrm{Real}({\mathbf{i}_{\mathrm{cov}}(r_i)})\\
            \mathrm{Imag}(\mathbf{i}_{\mathrm{cov}}(r_i))\\
            \mathrm{Phase}(\mathbf{i}_{\mathrm{cov}}(r_i))\\
        \end{pmatrix}.
\end{equation}

Computing these features for each range bin leads to the final DNN input image of 
shape $N_\mathrm{r}\times N_\theta \times N_{\mathrm{feat}}$.

In comparison, in a conventional FMCW radar signal processing pipeline, the beamforming 
is only applied to detections in the range-Doppler map. 
In this work, the beamforming is applied for each range bin to obtain image-like data. 
In this way, the DNN can also exploit the spatial relation between neighbouring range bins. 

\begin{figure*}
	\centerline{\includegraphics[width=0.9\textwidth]{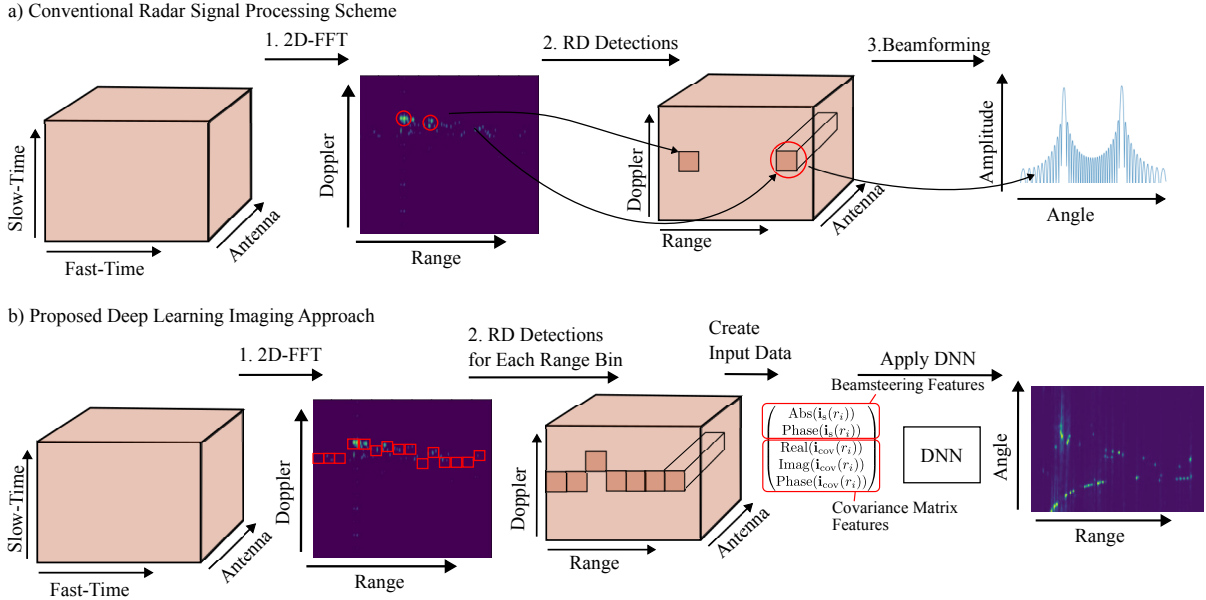}}
	\caption{
    The upper image describes the normal radar signal processing scheme.
    In the first step, the slow and fast-time is converted to range and Doppler information via a 2D-FFT. 
    Then, targets are detected in the range-Doppler map, and for each detection, the angle is
    estimated. In contrast, in the proposed approach b{)}, after computing the range-Doppler image for each virtual channel, 
        the maximum value of each range bin in Doppler direction is taken. Afterwards, different features are extracted from the range-channel data. 
        These are then used as an input for the DNN, resulting in high-resolution range-angle maps.} 
	\label{fig:SignalProcessingScheme}
\end{figure*}

The complete signal processing workflow starting from the raw, 3D radar cube is depicted in 
Fig.~\ref{fig:SignalProcessingScheme} b{)} and compared to 
the common signal processing in Fig.~\ref{fig:SignalProcessingScheme} a{)}.

\subsection{GROUND TRUTH DATA PROCESSING}\label{subsec:GroundTruthData}

The generation of the ground truth signal is described in detail in our earlier work~\cite{lit:schussler2022deep}. 
It is based on a matched filter approach and resembles the description in~\cite{lit:ahmed2012advanced}.
The advantages of this technique are that this reconstruction approach is also valid for targets that are not
located in the far field. This is especially relevant for the deployed large array. 
Also, using matched filters typically results in a better resolution with a lower sidelobe level. These advantages 
come at the price of a longer reconstruction runtime. 
However, since this reconstruction technique is only used to generate ground truth data, 
it does not have to be executed in real-time during the actual measurement process.
Doppler and range information are obtained in an identical way to the input data procedure.

Examples for input and ground truth signals are shown in Fig.~\ref{fig:ExampleImages} a{)}, 
along with virtual camera images of the simulated scene.

\subsection{EXTENSION FOR MULTIPLE DOPPLER DETECTIONS IN A SINGLE RANGE BIN}\label{subsec:ExtensionMoreDoppler}
Obviously, the proposed DNN is only able to process a range-angle image for a single Doppler detection in each range bin. In most cases this might be sufficient, as for example 
shown in~\cite{lit:major2019vehicle}, where the maximum projections on each dimension of the radar cube are utilized to create detections in an automotive radar scenario.
But in general, more than one Doppler detection for a single range bin should be considered. 
The proposed approach can be extended by running the same DNN multiple times for different Doppler detections in the following way:

\begin{enumerate}
    \item Create Doppler detections for each range bin and sort them by their strength
    \item Run the DNN for the strongest Doppler detection in each range bin
    \item Run the DNN for the second strongest Doppler detection. If no second strongest detection does exists take the strongest one.
    \item Repeat the procedure for the remaining Doppler detections.
\end{enumerate}

\section{DEEP LEARNING APPROACH}
In this section, the deployed DNN is briefly described and a description of the training data generation is given. 
\subsection{DEEP NEURAL NETWORK}

\begin{figure*}
	\centerline{\includegraphics[width=0.8\textwidth]{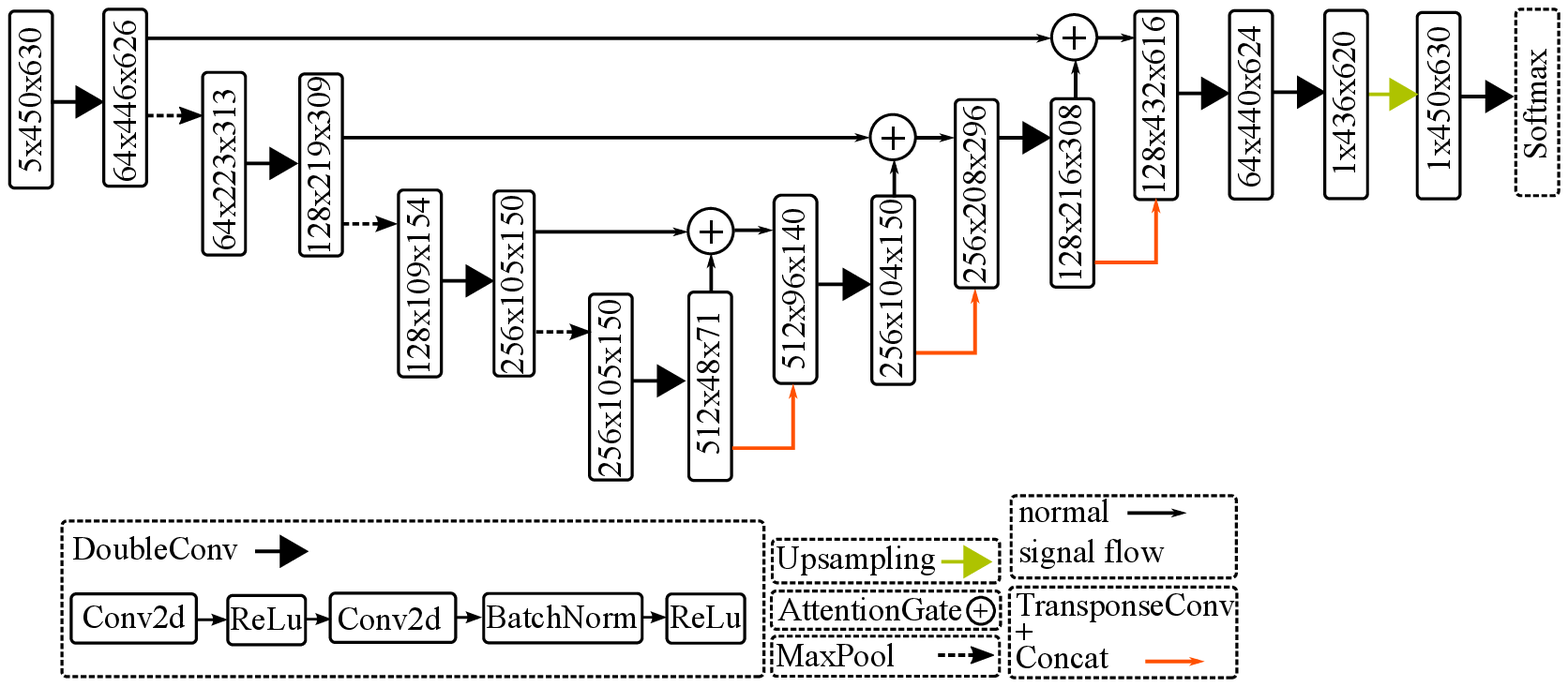}}
	\caption{The complete Attention U-Net architecture for the proposed problem, 
    slightly adapted from~\cite{lit:schussler2022deep}. Each input image consists of C channels with width W and height H, denoted as 
    $C\times W \times H$ in the scheme. For example, the input image consists of two channels
    with a width of 450 (representing the angular dimension) and a height of 630 (representing the range dimension).} 
	\label{fig:DNNScheme}
\end{figure*}

The DNN used for this approach is a conventional U-Net~\cite{lit:ronneberger2015u} augmented with an attention mechanism, 
adapted from~\cite{lit:oktay2018attention}. 
This network was chosen because it showed good results in previous work related to radar and especially to radar imaging tasks~\cite{lit:orr2021coherent, lit:schussler2022deep, lit:wei2018deep, lit:xu2020deep}.
The complete DNN is depicted in Fig.~\ref{fig:DNNScheme}, featuring the $N_{\mathrm{feat}}=5$ input channels as described in the previous section.
A softmax activation function can be added as the last activation function for a classification task. 
It should be mentioned, that the same network architecture as in our previous work~\cite{lit:schussler2022deep} was adapted.
The difference in the performance is achieved by choosing more suitable input features as described in the previous section.

In this setup, there are two ways to formulate the proposed problem. 
Firstly, it can be stated as a detection problem.
In the machine learning context, this is often described as a binary classification problem, 
where the decision has to be made if a target is present at a specific pixel or not.
Secondly, the problem can be viewed as a (potentially ill-posed) inverse problem. The first problem formulation 
is almost always chosen for the DoA deep learning approaches mentioned in Sec.~\ref{sec:RelatedWork}. 

The second problem formulation is very common in compressed sensing and 
in the deep learning image reconstruction methods mentioned in Sec.~\ref{sec:RelatedWork}.

Depending on how the problem is formulated, a different preprocessing of the ground truth data, 
a different loss function, and different activation functions are chosen.
If the problem is formulated as a detection problem, 
every pixel in the output image is set to either zero or one depending on whether a detection is present or not. 
Furthermore, the binary cross-entropy (BCE) loss is commonly 
chosen as the loss function. Other possible loss functions for imbalanced class distributions, 
such as focal loss~\cite{lit:lin2017focal} have also been deployed for DoA estimation~\cite{lit:fuchs2022machine}.

Nevertheless, if the problem is stated as an inverse problem, 
the L2-Norm will typically be chosen with an optional regularization term to incorporate prior information. In compressed sensing, the signal 
is often expected to be sparse, and therefore, the L1-Norm on the estimated signal itself is added.
Table~\ref{tab:ProblemFormulation} summarizes the differences of both problem formulations for the proposed approach. 
Here, the variable $\mathrm{X}_{\mathrm{est}}$ describes the estimated radar image
and $\mathrm{Y}_{\mathrm{tru}}$ describes the ground truth radar image.
For the classification tasks, each pixel of the simulated ground truth 
range-angle image is set to either zero or one.

The preprocessing in the regression task allows for a scaling of the ground truth image with the scaling factor $\beta$.
This is required if the pixels are too close to the background, 
since the DNN might optimize them to a minimum with all pixels being set to zero. 
Scaling the values to from zero may lead 
the learning process to another more suitable minimum, where the foreground pixels are detected properly.
In our approach, $\beta = 10$ has generated satisfactory results. We did not conduct a exhaustive grid search for this parameter, 
but setting the value to $\beta = 1$ resulted in a training process that did not converge.

In the loss term, the sparsity assumption of the output image can be adjusted via the parameter $\alpha$.
Setting this parameter to zero already leads to good results, since the ground truth data initially encodes the sparsity information.
In our case, even relatively small values ($\alpha = 0.1$) degraded the results slightly.
However, in scenarios with less training data, setting $\alpha$ to a value different to zero could be beneficial.
Also, the use of the logarithmized input magnitude data, instead of the linear magnitude, 
improves the training performance in this scenario.

Both problem formulations were thoroughly investigated, but the training for the regression 
task achieved better results in respect to detection performance and aliasing artifacts. This is because the sidelobes could not be fully removed in the ground truth data and were therefore 
equally weighted as real targets in the classification task. 
Also, the amplitude in the ground truth image still contains useful information 
that is completely removed when each target is set to a fixed value of one.

\subsection{TRAINING DATA GENERATION}
Accurate generation of sufficient training data is crucial for the performance of the DNN. 
For this, two approaches are pursued.

The first approach uses data generated by realistic radar simulations in a detailed virtual world based 
on the approach presented in~\cite{lit:schussler2021realistic}.
The procedure is identical to that in~\cite{lit:schussler2022deep}, where the virtual sensor was placed randomly at predefined positions 
on a virtual map imported from the open source simulator CARLA~\cite{lit:dosovitskiy2017carla}.
Additionally, for each virtual measurement, the position of the virtual sensor was moved in a random 
direction with a random velocity of between 0 and 6~$\frac{\textrm{m}}{\textrm{s}}$.
In this way, 13 000 virtual measurements were generated. 

However, although these data accurately resembles real-world measurements, it does not contain 
enough examples with closely spaced targets in the same range bin. 
Consequently, this dataset does not sufficiently challenge the DNN 
to produce high-resolution output but mainly reduces noise and sidelobes.

Therefore, as a second approach for the generation of training data, 
we also simulated 5500 synthetic datasets composed of closely separated point-like targets.
This synthetic dataset with point-like targets could not be used on its own to train the network. However, 
in combination with sufficient ray tracing simulation data, the results could be improved to a relevant extent.

\begin{figure}
	\centerline{\includegraphics[width=0.5\textwidth]{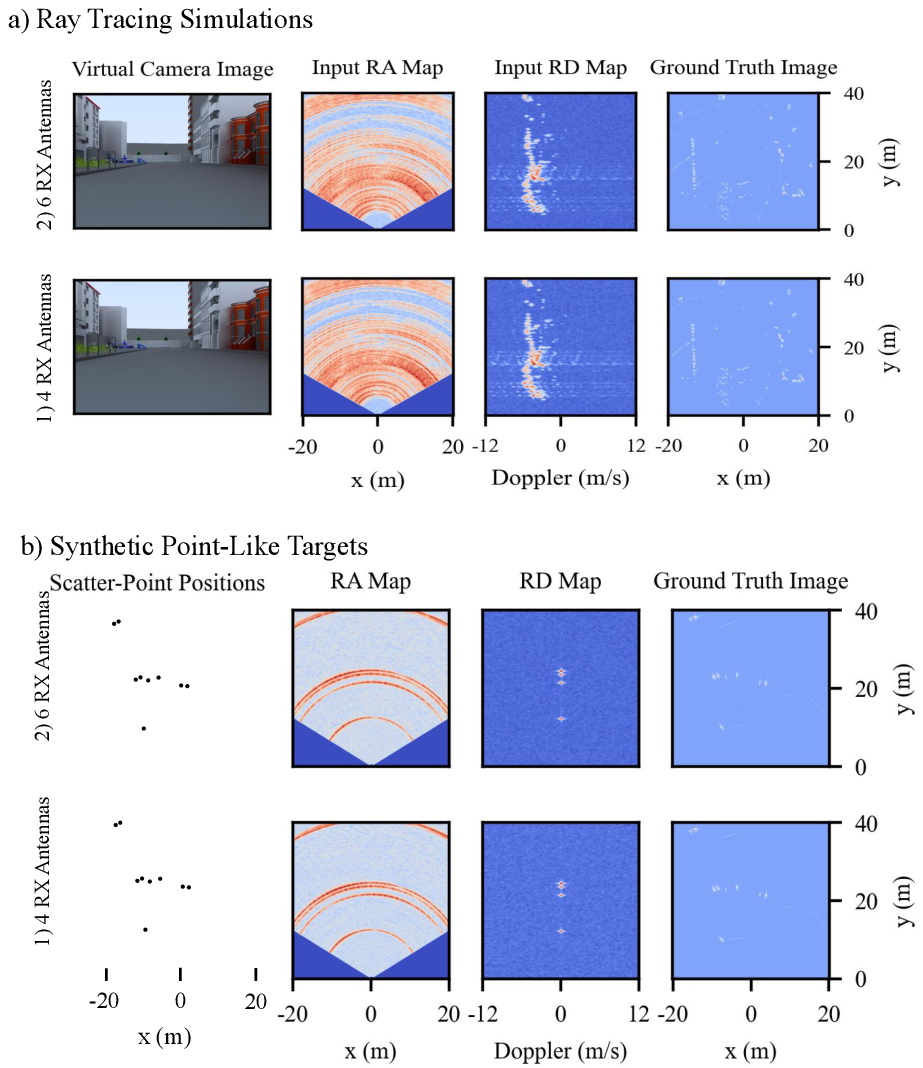}}
	\caption{The upper images in a{)} show example data for the two different antenna configurations 
            generated by ray tracing simulations. The image in the first column depicts
            a virtual camera image of the simulated scene, 
            and the second column image shows range-angle (RA) maps produced by the DaS beamforming technique. 
            Since the array is sparse, sidelobes dominate and no angular information can be perceived by the human eye. 
            The range-Doppler (RD) map is depicted in the third column image, and the ground truth image, based on the ULA with 
            256 elements, is presented in the right column.
            The lower images in b{)} show the same information, although without a simulated camera image. Here, synthetically generated
            static point-like targets were generated, which are closely spaced in angular direction.}
	\label{fig:ExampleImages}
\end{figure}

The two datasets were mixed together randomly, and 17~500 entries were used for training, 500 for validation, 
and 500 for test purposes.
Again, examples for both generated input images are shown in Fig.~\ref{fig:ExampleImages}.

\begin{table*}[ht!]
    \begin{center}
    \caption{Problem formulations}
    \label{tab:ProblemFormulation}
    \begin{tabular}{|c|c|c|}
    \hline
    & Binary Classification Task (Detection) &  Regression Task\\
    \hline
    Loss & 
    $l_{BCE} = \mathrm{mean}[-\mathbf{Y_{tru}} \cdot \log(\mathbf{X_{est}}) + (\mathbf{1}-\mathbf{Y_{tru}})\cdot \log(\mathbf{1}-\mathbf{X_{est}}) ]$
    & $l_{MSE} = \mathrm{mean}[||\mathbf{X}_{\textrm{est}} - \mathbf{X}_{\textrm{tru}}||_2 + \alpha||\mathbf{X}_{\textrm{est}}||_1]$\\
    \hline
    Pre-processing & 
    $\mathbf{Y_{tru}}(x,y)=
    \begin{cases}
    1, \textrm{ for } \mathbf{\hat{Y}_{tru}}(x,y) > 0\\
    0, \textrm{ otherwise }\\
    \end{cases}$
    &
    $\mathbf{Y_{\mathrm{tru}}} = \beta \mathbf{\hat{Y}_{\mathrm{tru}}} $\\
    
    \hline
    Last layer & Softmax activation function & No activation function \\
    \hline 
    \end{tabular}
    \end{center}
\end{table*}

\section{RESULTS}
In this section, the proposed DNN technique is compared to a 
common convolutional neural network (CNN)-based DoA algorithm trained 
on the covariance matrix of the proposed sparse antenna array configurations.\footnote[1]{Our DNN approach and the conventional CNN DoA estimator are both convolutional networks, but for the sake of simplicity, the former is abbreviated as DNN and the other as CNN in the following.}
This CNN DoA estimator proposed in~\cite{lit:papageorgiou2021deep} was chosen as the baseline,
since it showed very good and robust performance in a comparison conducted in~\cite{lit:fuchs2022machine}. 
The training and pre-processing is described in more detail in subsection~\ref{subsec:ReferenceImplementation}.

Furthermore, a conventional DaS beamforming technique using a hanning window, and the MUSIC algorithm were used for comparison.
For the MUSIC approach, the number of targets were estimated using the AIC criterion~\cite{lit:akaike1974new} 
with exact formulas taken from~\cite{lit:madisetti1997digital}.

\subsection{REFERENCE IMPLEMENTATION}\label{subsec:ReferenceImplementation}
The baseline DoA estimator from \cite{lit:papageorgiou2021deep} relies on the normalized covariance matrix $\mathbf{\Sigma}$
generated by the antenna channels as input:
\begin{equation}\label{equ:CovMat}
    \mathbf{\Sigma} = \frac{\mathbf{x} \mathbf{x}^\textrm{H}}{||\mathbf{x} \mathbf{x}^\textrm{H}||_2}.
\end{equation}
Here, $\mathbf{x}$ denotes the MIMO antenna channels resorted as for the 
virtual array depicted in Fig.~\ref{fig:AntennaConfig}.
Compared to the proposed approach, the beamforming technique takes only a single 
antenna channel line as input and no additional input is required.
Before being passed on to the CNN, the complex covariance matrix 
is split and stacked to a tensor of the dimension $3\times \mathrm{N_v} \times \mathrm{N_v}$ according to the following conversion:
\begin{equation}\label{equ:CovMatSplit}
\mathbf{\Sigma}_{\mathbf{T}} =  
    \begin{pmatrix}
        \mathrm{Re}({\mathbf{\Sigma}})\\
        \mathrm{Im}(\mathbf{\Sigma})\\
        \mathrm{\arg}(\mathbf{\Sigma})
    \end{pmatrix}.
\end{equation}

The CNN is loosely trained following the instructions from~\cite{lit:fuchs2022machine}, 
although adjusted to the fewer targets and a higher SNR, expected in our test data. The processing steps are as follows:

\begin{enumerate}
  \item Generate a random number of targets N (between 1 and 3).
  \item Assign an amplitude to each target, which is chosen from a specific range.
  \item Create an IF signal with a single time-snapshot, as described above.
  \item Add complex normally distributed noise to the signal with a random SNR value 
         from a specific predefined range (from 3 to \SI{30}{dB} in this work).
  
\end{enumerate}
In this case, the SNR is defined as the ratio between the weakest scatterer and the noise power, 
as adapted from~\cite{lit:wang2016coarrays}. The data were generated online, and 
the neural network was trained on 20 million samples.

\subsection{EVALUATION OF RESOLUTION AND POINT-SPREAD-FUNCTION}
To evaluate the proposed DNN approach, several real-world measurements were conducted.

\begin{figure*}
	\centerline{\includegraphics[width=1.0\textwidth]{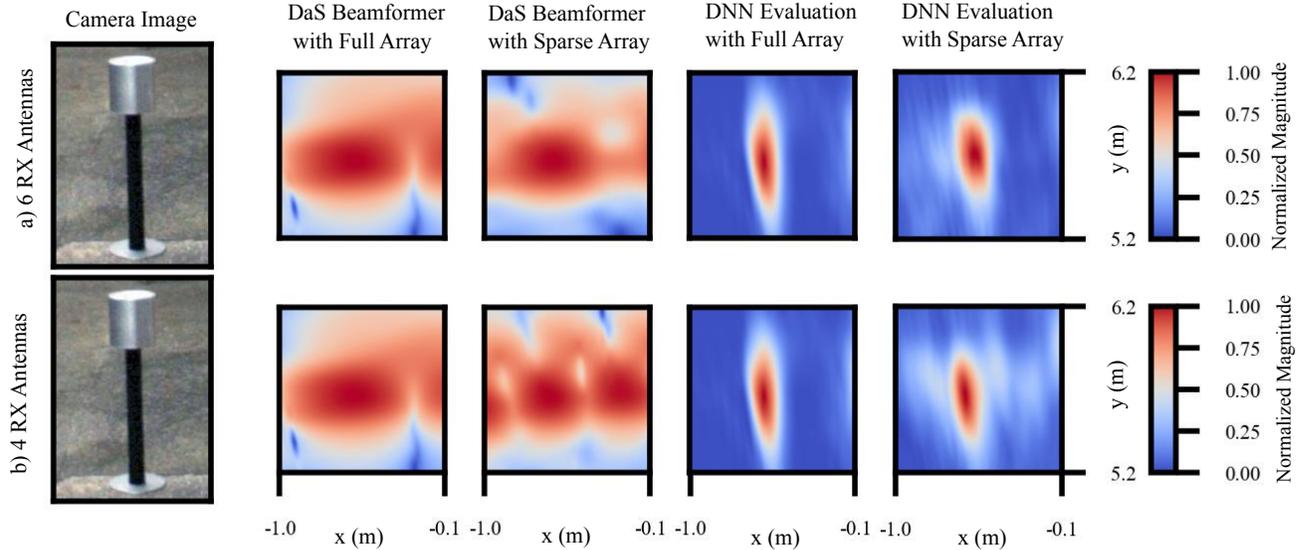}}
	\caption{PSF of the proposed DNN approach compared to a conventional DaS beamforming. 
    The first two columns show the PSF of a conventional DaS beamforming, the first (a{)} for a 
    fully occupied array and the second (b{)} for the sparse array configuration. 
    The same is shown for the DNN approach in the right two columns.}
	\label{fig:PSF}
\end{figure*}

In the first test, a single metal cylinder at a distance of approximately \SI{5.7}{m} was observed and the 
corresponding point spread function (PSF) of the DNN result compared to the conventional DaS beamforming approach.
The results are presented in Fig.~\ref{fig:PSF}.
Clearly, the DNN not only led to a much sharper PSF but also removed aliasing artifacts almost completely.

\begin{figure*}
	\centerline{\includegraphics[width=1.0\textwidth]{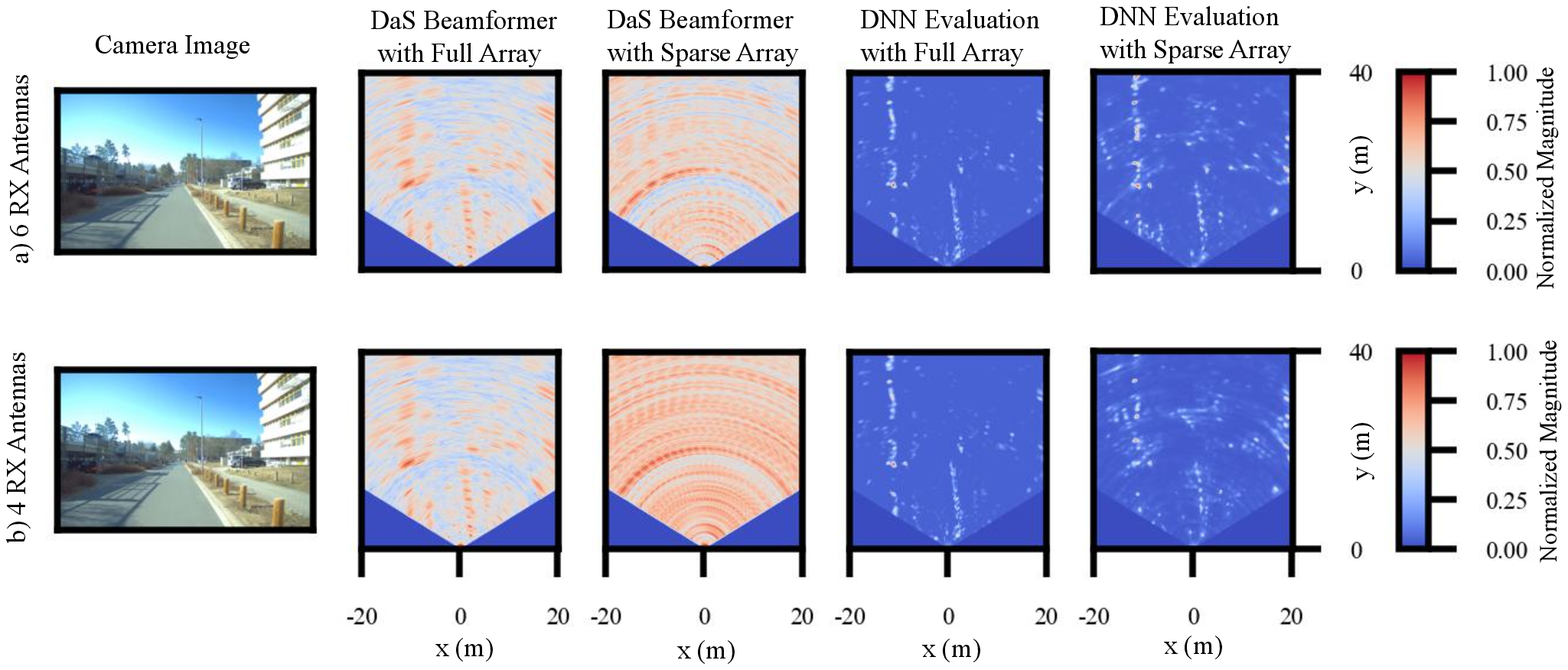}}
	\caption{Evaluation of the proposed DNN approach on a realistic automotive radar scene. 
    The first two columns show the results using a conventional DaS beamforming, the first (a{)}) for a 
    fully occupied array and the second (b{)}) for the sparse array configuration. 
    The same is shown for the DNN approach in the right two columns.}
	\label{fig:RealMeasurement}
\end{figure*}

In the second test, the DNN results were compared to the DaS beamforming approach in realistic 
automotive scenarios, shown in Fig.~\ref{fig:RealMeasurement}.
Similar to the previous results, the DNN results were almost aliasing-free, even for 4 RX antennas, 
which are significantly more challenging.

As in comparison, the conventional DaS beamforming technique was not capable of evaluating 
useful angular information due to massive aliasing artifacts.

\begin{figure*}
	\centerline{\includegraphics[width=1.0\textwidth]{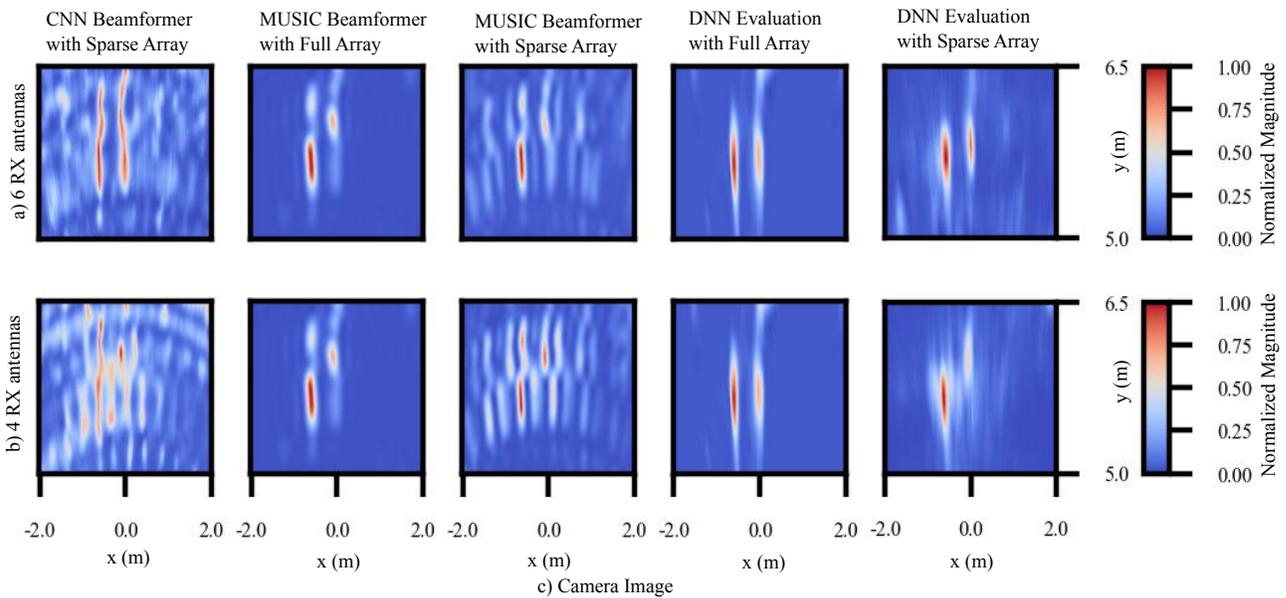}}
	\caption{Evaluation of the proposed DNN approach compared to a conventional CNN approach as well as a MUSIC 
        beamformer in a scenario testing the angular resolution. 
        The first column shows the results of a state-of-the-art CNN for the sparse array. 
        Since the input covariance matrix is too large, the results could not be generated for the full array configuration.
        The second and third column depict the imaging results using the super-resolution algorithm MUSIC 
        for a fully occupied ULA and the sparse array configuration. The fourth and fifth column show the results of
        the DNN approach, again for the fully occupied and the sparse array, respectively.
        a{)} presents the results for the sparse array from Fig. 4 using 6 RX antennas, and b{)} gives 
        the results for the second configuration using 4 RX antennas. 
        In c{)}, the camera image of the two cylinders acting as point-like targets is shown.}
	\label{fig:ResolutionEval}
\end{figure*}

In the third test, the angular resolution of the implemented DNN was evaluated and compared to the MUSIC algorithm and 
a state-of-the-art CNN beamforming approach introduced before. The results are depicted in Fig.~\ref{fig:ResolutionEval}. 
The evaluation of the CNN approach for a full array was not possible, since the full covariance matrix required for this approach was too large. 
Comparing the results for the sparse array configurations, the DNN approach outperformed the other algorithms in terms of sidelobe 
suppression and resolution. In particular, for the array configuration with only 4 RX antennas depicted in Fig.~\ref{fig:ResolutionEval} b{)}, the DNN 
is still capable of separating both targets. 
Due to the high aliasing, this is not feasible with the other approaches presented.

\subsection{EVALUATION OF DETECTION AND FALSE-ALARM PROBABILITY}

As a fourth test, the proposed DNN approach was evaluated in terms of precision, detection rate, and false alarm rate.
These metrics are briefly described below.
\begin{enumerate}
    \item The detection rate $P_\mathrm{D}$, also called \emph{recall}:
    \begin{equation}
        P_{\mathrm{D}} = \frac{\textrm{TP}}{\textrm{TP} + \textrm{FN}} \textrm{, }
    \end{equation}
    \item The false alarm rate $P_{\mathrm{FA}}$:
    \begin{equation}
        P_{\mathrm{FA}} = \frac{\textrm{FP}}{\textrm{TN} + \textrm{FP}} \textrm{,}
    \end{equation}
    \item The precision, as the ratio between correct detections and all detections:
    \begin{equation}
        \mathrm{Precision} = \frac{\textrm{TP}}{\textrm{TP} + \textrm{FP}} \textrm{.}
    \end{equation}
\end{enumerate}

Here, TP and TN denote true positives and true negatives, respectively.
FP and FN describe false positives and false negatives, respectively. 
The approach was again applied to simulated data, which were neither used for training or validation. 
The test set consisted of 50 images, in which again each range bin with at least one detection 
in the ground truth data was considered. 

The evaluation was conducted for each simulated dataset in the following way.

\begin{enumerate}
    \item Create the range-Channel image and extract the necessary features for the deployed DoA estimators. 
    \item Apply the DNN approach on the complete test image and store the resulting image.
    \item Iterate over all range bins and apply the beamforming algorithms (CNN and MUSIC). For the DNN approach take the selected line from the already processed image.
    \item Apply a peak search algorithm for the currently selected range bin to find possible detections. As peak search algorithm, the {\tt{find\_peaks}} method 
    from the python library scipy\footnote[1]{https://docs.scipy.org} was used. 
    The sensitivity of the peak search was varied with the \emph{prominence} parameter, which according 
    to the documentation is \emph{a measure of how much a peak stands out from the surrounding baseline.}\footnote[2]{https://docs.scipy.org/doc/scipy/reference/generated/scipy.signal.peak
    \_prominences.html}. For more information about the algorithm and its parameters, the reader is refered to documentation of the algorithm.
    \item A detected peak is counted as successfully detected (TP) if 
    \begin{equation}
        p_{\mathrm{suc}} = 
        \begin{cases}
            \mathrm{TP}\textrm{, } | x_{\mathrm{est}} - x_{\mathrm{true}} | \leq d_{\mathrm{max}}\\
            \mathrm{FP}\textrm{, }\textrm{otherwise}\\
        \end{cases} \textrm{,}
    \end{equation}
    where, $x_{\mathrm{est}}$ denotes the pixel position of the estimated target, $x_{\mathrm{true}}$ is the correct target position, and
    $d_{\mathrm{max}}$ describes the maximum tolerated distance in pixels between $x_{\mathrm{est}}$ and $x_{\mathrm{true}}$. 
    In the proposed evaluation scenario, $d_{\mathrm{max}} = 2$ and $d_{\mathrm{max}} = 4$ was tested.  
    For the sake of simplicity, one estimated detection can be assigned to multiple ground truth detections. 
    A false negative (FN) detection was counted if the a peak was present in the ground truth signal, but was not detected by the beamforming algorithm.
    In case no peak was detected and was not present in the ground truth signal, a true negative (TN) was counted.
    
\end{enumerate}

Further, every spectrum assigned to the peak finding method was normalized to a range of 0 and 1. 
Since the CNN approach used for comparison in particular suffers from severe 
sidelobes, only peaks higher than 0.5 were counted. 
Even the proposed DNN beamforming did not considerably benefit from this additional threshold, this resulted in a fairer comparison.

\begin{figure*}
	\centerline{\includegraphics[width=0.95\textwidth]{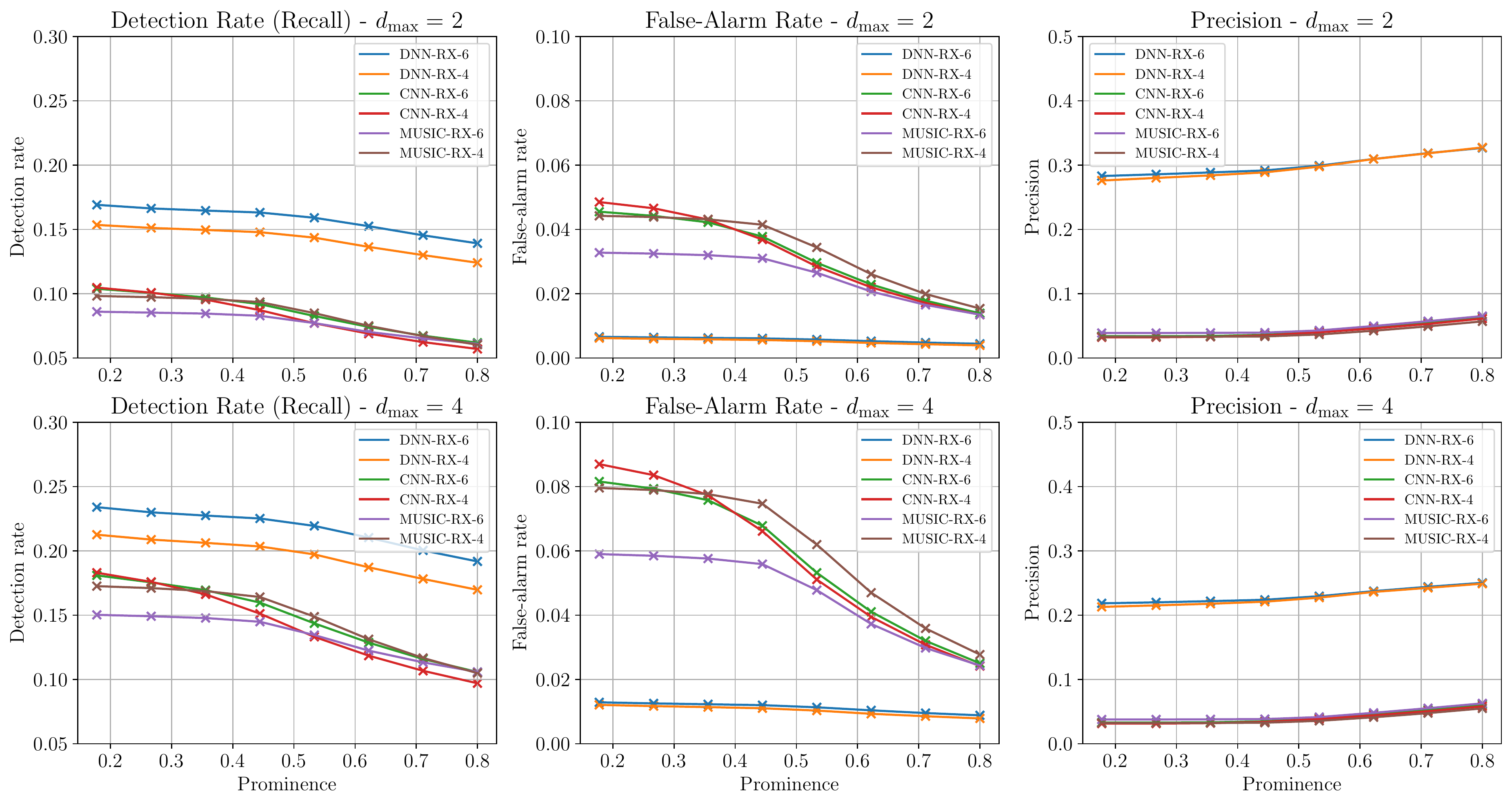}}
	\caption{Evaluation of three metrics for the comparison of the DNN approach with the conventional CNN method. 
    The columns indicate the specific metrics, which are detection rate, false alarm rate, and precision. 
    The rows indicate the performance, for the specific metric, of the investigated algorithms. 
    An estimated detection is counted if the distance to the ground truth detection is less than or equal 
    to the maximum distance $d_{\mathrm{max}}$ in the angular bins.} 
	\label{fig:Metrics}
\end{figure*}

The results are shown in Fig.~\ref{fig:Metrics}.
It is apparent that all algorithms exhibited a rather low detection rate.
This may be due to the fact that the ground truth image was generated 
with a much higher angular resolution, which prevented smaller arrays from resolving every target.  
It is nonetheless obvious that the proposed DNN approach outperformed the CNN method significantly. 
This is presumably due to the improved ratio between the mainlobe and the sidelobe magnitudes of the detected targets. Consequently, 
the peak search algorithm could more easily identify peaks as targets instead of noise or aliasing artifacts.
Interestingly, the MUSIC approach for only 6 RX antennas showed a slightly worse detection performance compared by using only 4 RX antennas.
This might be explained by the fact, that the aperture size for both antenna configurations is identical, but the MUSIC algorithm with only 4 antennas creates 
more sidelobes, which may be at the same position as real targets.

The same behavior could be observed for the false-alarm rate, which is considerably lower for the DNN approach compared to the MUSIC and CNN techniques. 
Even with an additional threshold during the peak search, the distortion from sidelobes was still very severe in the other approaches.

Contrary to the detection rate, the false-alarm rate of the MUSIC algorithm with 6 RX antennas is significantly better compared by using only 4 antennas.
This fits our previous explanation, that the detections for 4 RX antennas mainly stems from sidelobes, which also lead to large a amount of false detections resulting
in a high false-alarm rate.
The same argumentation can be applied to the precision metric, which is similar for the MUSIC with the CNN technique and the DNN approach showing the best results.
As expected, all beamforming algorithms exhibited a higher detection rate with increasing $d_{\mathrm{max}}$.

Interestingly, in this comparison, the difference between both DNN approaches was not very significant. 
This may be explained by the dataset used for comparison, which contained simulated 
but quite realistic radar images, where the task of differentiating two closely spaced targets did not happen often. 
Therefore, even though the DNN trained with 4 RX antennas 
had a lower resolution, it still showed a good sidelobe suppression, 
while maintaining a sufficient resolution for most of the tasks at hand.

\subsection{EVALUATION OF A DYNAMIC SCENE}
\begin{figure*}
	\centerline{\includegraphics[width=0.8\textwidth]{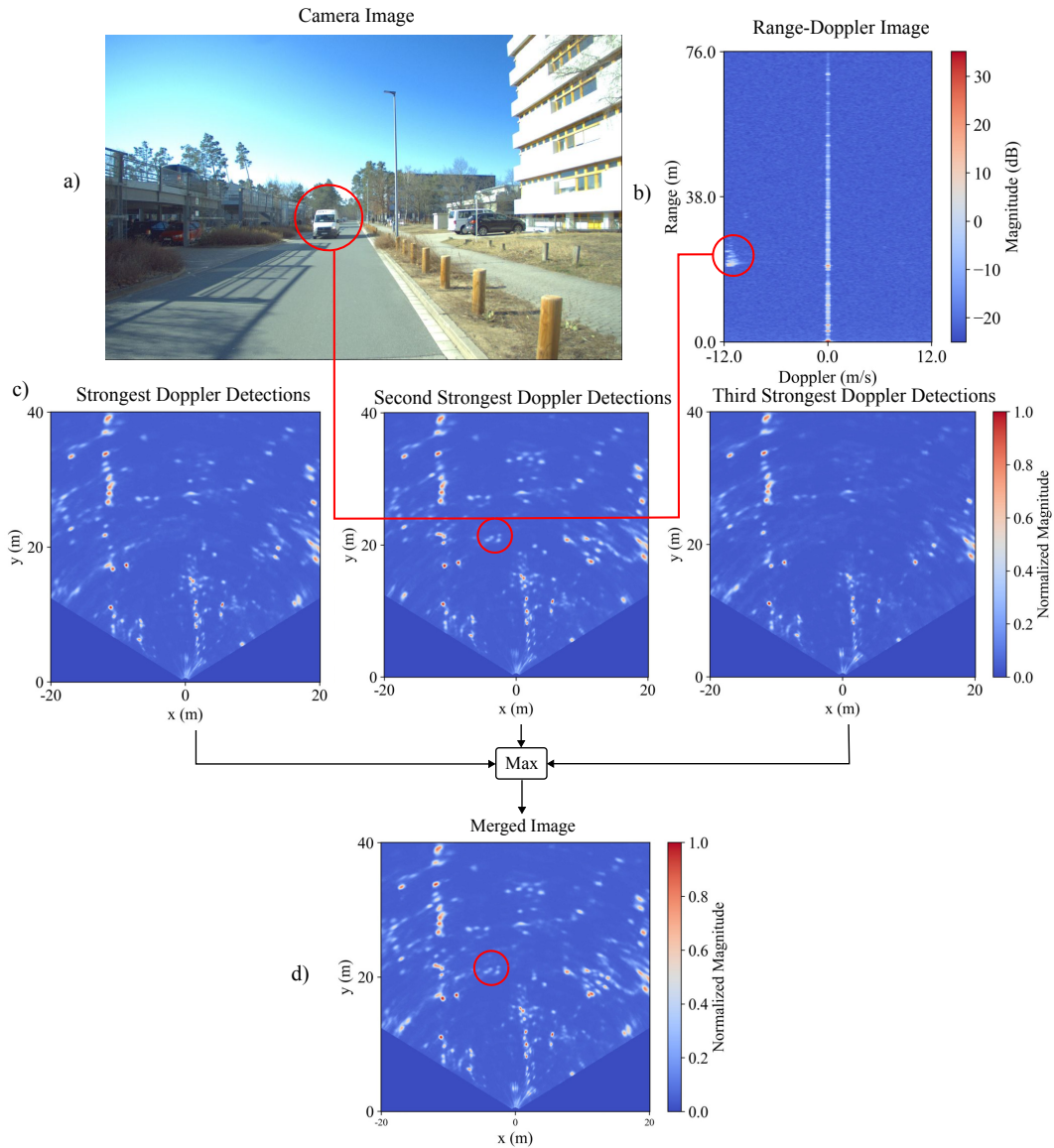}}
	\caption{Image a{)} shows the camera image with the approaching truck and image b{)} shows the corresponding range-Doppler image, in which the truck can
    clearly be separated in the Doppler direction. The generated Doppler detections are sorted and used to generate three range-angle images by applying the DNN. 
    All images are stacked onto each other and the maximum value for each pixel along these images is taken to generate a merged image.}
	\label{fig:DynamicScene}
\end{figure*}

Here, the approach described in Sec.~\ref{sec:SigProc}. \ref{subsec:ExtensionMoreDoppler} with the DNN trained for 6 RX antennas is evaluated on a dynamic scene, in which a truck is moving 
towards the radar unit. In this case, not only the maximum Doppler is chosen, but the peak search algorithm utilized in the previous section was used to 
obtain up to three Doppler peaks for each range bin. The result is depicted in Fig.~\ref{fig:DynamicScene}. The final image is created by stacking all three 
single images onto each other and taking the maximum projection along the new channel dimension, as shown in image d{)} in Fig.~\ref{fig:DynamicScene}. Alternatively, 
also detections from each single image could be collected and written in a common detection list.

This example illustrates, that evaluating multiple Doppler peaks is a reasonable measure, since only in the second image (Fig.~\ref{fig:DynamicScene} c)
the detections originating from the truck are clearly visible.

\section{CONCLUSION AND FUTURE WORK}
This work demonstrates that adequately preprocessed radar data from an extremely thinned aperture can be processed 
aliasing-free and with super-resolution using a DNN. For this, the DNN was specially 
trained with sufficient realistic simulation data obtained from advanced ray tracing simulations 
and with synthetically generated point-like targets. These targets were closely spaced to explicitly 
train the resolution capability of the DNN.
With the two evaluated sparse array configurations, it could be seen that the proposed 
approach outperformed even state-of-the-art machine learning DoA algorithms and super-resolutions techniques such as MUSIC.

In the following, we discuss possible reasons why this approach is beneficial 
in the proposed scenarios and make some suggestions for further improvements.

The first advantage of the proposed approach is that the input data do not consist of only a single covariance matrix 
or a single snapshot of the antenna array data.
Instead, the complete range-angle map and covariance information are given to the DNN, 
which can exploit specific spatial features encoded in the input data to improve the radar image result.
In addition, due to advanced ray tracing simulation, the training data used are much more realistic than simple 
point-like targets used in other approaches.

However, there are some possible areas for further improvement.
The proposed DNN is quite large, and it should be investigated 
if similar performance is possible with smaller architectures with fewer parameters. 

Furthermore, optimized array configurations should also be analyzed, since the virtual antenna array configurations investigated 
were limited by the MIMO configurations, and in this work, only the RX channels were reduced.
However, the fact that this approach performs well even on such unconventional 
array configurations opens up more flexibility to array design.

The evaluation of the proposed approach also revealed several promising aspects for future work.
Because this approach works well for the unconventional array configurations presented, 
it provides an unprecedented flexibility for the array design, which often suffers 
from strict limitations in terms of spatial constraints, power consumption, calibration issues, and raw data size.

Additionally, since the virtual antenna positions are comparably widely spaced, 
this approach is promising for the fusion and joint DoA estimation of distributed radar sensors.

Finally, also, other more recent neural network architectures, such as vision transformer~\cite{lit:dosovitskiy2020image, lit:liu2021swin}, which have been already applied successfully to radar
applications~\cite{lit:giroux2023t} could be investigated in respect of image quality and runtime performance in the future.

\section*{ACKNOWLEDGMENT}
The authors would like to thank the Symeo team from
indie Semiconductor (Jannis Groh and Javier Martinez) for their support with the radar system used for
tests.

\bibliographystyle{IEEEtran}
\bibliography{IEEEabrv, mybib}

\begin{IEEEbiography}[{\includegraphics[width=1in,height=1.25in,clip,keepaspectratio]{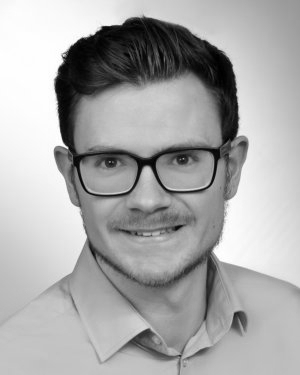}}]{CHRISTIAN SCHUESSLER}{\space}(Graduate Student Member, IEEE) 
    was born in Fürth, Germany, in 1991. He was awarded B.Eng and M.Eng. degrees in Electrical Engineering and Information Technology in 2014 and 2015, respectively, 
    from the Technische Hochschule Nürnberg Georg-Simon-Ohm, Nuremberg, Germany. 
    From 2015 to 2019 he worked as a software engineer in the algorithmic group of the Fraunhofer Center for X-Ray Development (EZRT) in the field of computer tomography. 
    He has been working toward his Ph.D. at the institute of
    Microwaves and Photonics (LHFT) at the Friedrich-Alexander-Universität Erlangen-Nürnberg, Erlangen, Germany since 2019. 
    His research focuses on radar simulation, signal processing, and machine learning, especially for large MIMO arrays in automotive applications.
\end{IEEEbiography}

\begin{IEEEbiography}[{\includegraphics[width=1in,height=1.25in,clip,keepaspectratio]{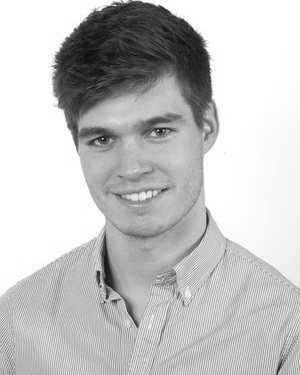}}]{MARCEL HOFFMANN}{\space}(Graduate Student Member, IEEE) 
    was born in Witten, Germany, in 1994. He received the B.Sc. and the M.Sc. degrees in 
    Electrical Engineering from the Friedrich-Alexander-Universität Erlangen-Nürnberg, Erlangen, Germany in 2016 and 2018, respectively. 
    In 2018, he joined the Institute of Microwaves and Photonics (LHFT) at the Friedrich-Alexander-Universität Erlangen-Nürnberg, 
    where he is currently working toward his Ph.D. His research focuses on radar signal processing and automotive radar applications including new SAR and SLAM approaches.
\end{IEEEbiography}

\begin{IEEEbiography}[{\includegraphics[width=1in,height=1.25in,clip,keepaspectratio]{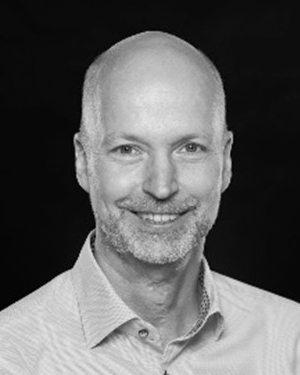}}]{MARTIN VOSSIEK}{\space}(Fellow, IEEE) 
    received the Ph.D. degree from Ruhr-Universität Bochum,
	Bochum, Germany, in 1996.
    In 1996, he joined Siemens Corporate Technology, Munich, Germany,
    where he was the Head of the Microwave Systems Group from 2000 to 2003. Since 2003, 
    he has been a full professor with Clausthal University, Clausthal-Zellerfeld, Germany. 
    Since 2011, he has been the Chair of the Institute of Microwaves and Photonics (LHFT), 
    Friedrich-Alexander-Universität Erlangen-Nürnberg (FAU), Erlangen, Germany. He has authored or coauthored more than 350 articles. 
    His research has led to more than 100 granted patents. 
    His current research interests include radar, microwave systems, wave-based imaging, transponders, RF identification, communication, and wireless sensor and locating systems. 
    Dr.~Martin Vossiek is a member of the German National Academy of Science and Engineering (acatech) and of the German Research Foundation (DFG) Review Board. 
    He is a member of the IEEE Microwave Theory and Technology (MTT) Technical Committees for MTT-24 Microwave/mm-wave Radar, Sensing, and Array Systems; 
    MTT-27 Connected and Autonomous Systems (as founding chair); and MTT-29 Microwave Aerospace Systems. 
    He also serves on the advisory board of the IEEE CRFID Technical Committee on Motion Capture \& Localization. 
    Dr.~Martin Vossiek has received numerous best paper prizes and other awards. In 2019, he was awarded the Microwave Application Award by the IEEE MTT Society (MTT-S) 
    for Pioneering Research in Wireless Local Positioning Systems. 
    Dr.~Vossiek has been a member of organizing committees and technical program committees for many international conferences and has served on the review boards of numerous technical journals. 
    From 2013 to 2019, he was an associate editor for the IEEE Transactions on Microwave Theory and Techniques. Since October 2022, he has been the associate Editor-in-Chief for IEEE Transactions on Radar Systems.
\end{IEEEbiography}

\end{document}